\documentstyle[12pt,aps]{revtex}
\begin{document}
\draft
\title{\large {\bf The nuclear shell effects near the r-process path
in the relativistic Hartree-Bogoliubov theory}}
\author{Madan M. Sharma and Ameena R. Farhan}
\address{Physics Department, Kuwait University, Kuwait 13060}
\date{\today}
\maketitle

\begin{abstract}
We have investigated the evolution of the shell structure of nuclei 
in going from the r-process path to the neutron drip line within the 
framework of the Relativistic Hartree-Bogoliubov (RHB) theory.
By introducing the quartic self-coupling of $\omega$ meson in the RHB 
theory in addition to the non-linear scalar coupling of $\sigma$ 
meson, we reproduce the available data on the shell effects
about the waiting-point nucleus $^{80}$Zn. With this approach, it is 
shown that the shell effects at $N=82$ in the inaccessible region of the
r-process path become milder as compared to the Lagrangian with the
scalar self-coupling only. However, the shell effects remain stronger
as compared to the quenching exhibited by the HFB+SkP approach. It is 
also shown that in reaching out to the extreme point at the neutron drip 
line, a terminal situation arises where the shell structure at the magic
number is washed out significantly.

\end{abstract}


\section {Introduction}

A knowledge of the shell effects near the r-process path is important to 
discerning astrophysical scenario of nucleosynthesis \cite{Kratz.93}. 
The question whether the shell effects near the drip lines are strong 
or do quench has become crucial to understanding the nucleosynthesis
of heavy nuclei. The $N=82$ nuclei at the r-process path are assumed to 
play a significant role in providing nuclear abundances about $A \sim 130$.
Since nuclei contributing to this peak are extremely neutron-rich
and are not accessible experimentally, it has not been possible to
ascertain the nature of the shell effects in the vicinity of the 
r-process path. Due to the lack of experimental data, there prevail 
conflicting view points \cite{SLHR.94,DHNA.94} on the strength 
of the shell effects near the r-process path. 

In ref. \cite{SLHR.94} the shell effects in the region of the r-process 
path at $N=82$ were studied on the basis of the Relativistic Mean-Field 
(RMF) theory using a successful nuclear interaction. It was predicted 
\cite{SLHR.94} that the shell effects at $N=82$ in Zr nuclei are strong. 
In contrast, the shell effects were suggested \cite{DHNA.94} to be 
quenched on the basis of the Hartree-Fock-Bogoliubov (HFB) calculations 
using the density-dependent Skyrme force SkP. In the absence of experimental 
data on the shell effects for $N=82$ nuclei near the r-process path, however, 
it is difficult to verify either of the two predictions. In view of this, 
we first attempt to reproduce the available data on the shell effects 
in the known regions and then extrapolate the method to the inaccessible 
region. 

In this paper, we examine how the shell effects evolve with isospin
in the region of the astrophysically important magic number $N=82$ near 
the r-process path. Using the experimental data available on the 
waiting-point nucleus $^{80}$Zn ($N=50$), we explore the shell 
effects near the r-process path in the framework of the Relativistic
Hartree-Bogoliubov (RHB) theory with the self-consistent
finite-range pairing. In order to improve the description of the shell 
effects, we have also introduced the vector self-coupling of 
$\omega$ meson in addition to the non-linear scalar coupling of the 
$\sigma$-meson in the RHB theory. It has been shown that with this approach
the high-precision experimental data on the shell effects
about the stability line can be reproduced well \cite{SFM.00}. This
forms our basis to be able to predict as to how the shell effects 
evolve in the unknown region.   

The shell effects are known to manifest strongly in terms of the magic 
numbers. This is signified by a prominent kink about the major magic 
numbers as can be seen vividly in the 2-neutron separation energies 
($S_{2n}$) all over the periodic table \cite{Bor.93}. This is a
demonstration of the existence of large shell gaps at the magic 
numbers in nuclei about the stability line.
Here the spin-orbit interaction is pivotal in the creation of the magic 
numbers \cite{MGM.55}. In the RMF theory \cite{Serot.86} the spin-orbit 
term arises naturally as a consequence of the Dirac-Lorentz structure of 
nucleons. This has shown much usefulness in explaining properties which 
involve shell effects such as anomalous isotope shifts in stable nuclei 
\cite{SLR.93}. The form of the spin-orbit interaction in the RMF theory 
has been found to be advantageous over that in the non-relativistic 
approaches \cite{SLKR.94}. 

For nuclei near the drip line, a coupling to the continuum is required.
A self-consistent  treatment of pairing is also desirable. The framework
of the RHB theory provides an appropriate tool to include both these
features. Thus, in the RHB theory the advantage of a relativistic
description of the RMF approach in the Hartree channel is combined 
with that of a self-consistent finite-range pairing force. We provide a 
short summary of the RHB theory in the next Section. In section III we 
give the details of the calculations. The ensuing results are presented 
and discussed in Section IV. Here a detailed comparison of the shell 
effects using different forces and approaches is made. The single-particle 
levels are used to illustrate the nature of the shell effects in the 
different approaches. The last section concludes the results of the paper. 

\section{The Relativistic Hartree-Bogoliubov Theory}

The RMF Lagrangian which describes the nucleons as Dirac spinors 
moving in meson fields is given by \cite{Serot.86}
\begin{eqnarray}
{\cal L}&=& \bar\psi \left( \rlap{/}p - g_\omega\rlap{/}\omega -
g_\rho\rlap{/}\vec\rho\vec\tau - \frac{1}{2}e(1 - \tau_3)\rlap{\,/}A -
g_\sigma\sigma - M_N\right)\psi\nonumber\\
&&+\frac{1}{2}\partial_\mu\sigma\partial^\mu\sigma-U(\sigma)
-\frac{1}{4}\Omega_{\mu\nu}\Omega^{\mu\nu}+ \frac{1}{2}
m^2_\omega\omega_\mu\omega^\mu\\ &&+\frac{1}{2}g_4(\omega_\mu\omega^\mu)^2
-\frac{1}{4}\vec R_{\mu\nu}\vec R^{\mu\nu}+
\frac{1}{2} m^2_\rho\vec\rho_\mu\vec\rho^\mu -\frac{1}{4}F_{\mu\nu}F^{\mu\nu}
\nonumber
\end{eqnarray}
where $M_N$ is the bare nucleon mass and $\psi$ is its Dirac spinor. In 
addition, we have the scalar meson ($\sigma$), isoscalar vector meson 
($\omega$), isovector vector meson ($\rho$) and the electromagnetic field 
$A^\mu$, with the masses $m_\sigma$, $m_\omega$ and $m_\rho$ and the coupling 
constants $g_\sigma$, $g_\omega$, and $g_\rho$, respectively. The field 
tensors for the vector mesons are given as 
$\Omega_{\mu\nu}=\partial_\mu\omega_\nu-\partial_\nu\omega_\mu$ 
and by similar expressions for the $\rho$-meson and 
the photon. For a realistic description of nuclear properties a nonlinear
self-coupling $U(\sigma) = \frac{1}{2} m^2_\sigma \sigma^2_{} +
\frac{1}{3}g_2\sigma^3_{} + \frac{1}{4}g_3\sigma^4_{}$ for  
$\sigma$-mesons has become standard. We have added the non-linear vector 
self-coupling of $\omega$-meson \cite{Bod.91}, which is represented by the 
coupling constant $g_4$. 

Using Green's function techniques \cite{Go.58} it has been shown in ref.
\cite{KR.91} that a relativistic Hartree-Bogoliubov theory can be
implemented using such a Lagrangian. Neglecting retardation effects one 
obtains a relativistic Dirac-Hartree-Bogoliubov (RHB) equations
\begin{equation}
\left(\begin{array}{cc} h & \Delta \\ -\Delta^* & -h^* \end{array}\right)
\left(\begin{array}{r} U \\ V\end{array}\right)_k~=~
E_k\,\left(\begin{array}{r} U \\ V\end{array}\right)_k,
\label{RHB} 
\end{equation}
where $E_k$ are the quasiparticle energies and the coefficients $U_k$ and 
$V_k$ are four-dimensional Dirac spinors normalized as
\begin{equation}
\int ( U^+_k U^{}_{k'}~+~V^+_kV^{}_{k'}\, ) d^3r~=~\delta_{kk'}.
\end{equation} 
The average field
\begin{equation}
h~=~\mbox{\boldmath $\alpha p$}~+~g_\omega\omega~+~ \beta(M+g_\sigma
\sigma)~-~\lambda
\label{h-field}
\end{equation}
contains the chemical potential $\lambda$ which is adjusted to the proper 
particle number. The meson fields $\sigma$ and $\omega$ are determined 
self-consistently from the Klein Gordon equations:
\begin{eqnarray}
\left\{-\Delta+m^2_\sigma\right\}\sigma&=& -g_\sigma \rho_s~-~g_2
\sigma^2~-~g_3\sigma^3,\\ 
\left\{-\Delta+m^2_\omega\right\}\omega&=& g_\omega \rho_v~+~g_4\omega^3,
\label{KG}
\end{eqnarray}
with the scalar density $\rho_s=\sum_k \bar V^{}_kV^{}_k$ and the baryon
density $\rho_v=\sum_k V^+_kV^{}_k$.  The sum on $k$ runs only over all the
particle states in the {\it no-sea} approximation. The pairing field 
$\Delta$ in Eq. (\ref{RHB}) is given by
\begin{equation}
\Delta_{ab}~=~\frac{1}{2}\sum_{cd} V^{pp}_{abcd} \kappa_{cd},
\label{gap}
\end{equation}
where $V^{pp}_{abcd}$ are the matrix elements of a general two-body
pairing interaction, and the pairing tensor is defined as
\begin{equation}
{\bf\kappa}_{cd}({\bf r},{\bf r}') = 
\sum_{E_k>0} U_{ck}^*({\bf r})V_{dk}({\bf r}').
\end{equation}
The RHB equations (\ref{RHB}) are a set of four coupled 
integro-differential equations for the Dirac spinors $U(r)$ and 
$V(r)$ which are obtained self-consistently. The solution of the 
Dirac-Hartree-Bogoliubov eigenvalue equations and the meson-field
equations determines the nuclear ground state.

Nuclei which are known to show strong pairing correlations
are treated appropriately within the framework of the
RHB approach. This is especially important when the pairing 
correlations in the middle of a shell become significant. 
On the other hand, the pairing correlations for nuclei 2 neutrons 
less or more than a magic number can be expected to be similar
in the RHB and the BCS approach. This is due to the reason that 
the pairing correlations in such nuclei are reduced to a minimal 
level. However, in the present work, we have performed the RHB calculations
for all nuclei. For the pairing channel, we have taken the finite-range 
Gogny force D1S. It is known to represent the pairing properties of a 
large number of finite nuclei appropriately 
\cite{Berger.36}.

The RHB approach becomes important while dealing with nuclei 
far away from the stability line and in particular for those in the 
vicinity of the drip lines where the Fermi level is usually very
close to the continuum. The RHB theory which is based upon the 
quasi-particle scheme employing Bogoliubov transformations takes 
into  account the coupling between the bound states and the
states in the continuum. The pairing and the coupling to the 
continuum can be important in nuclei with a pair of particles
above the shell closure in determining the shell effects especially
as the Fermi surface is very close to the continuum. Therefore, the RHB 
theory with the self-consistent pairing serves as an ideal scheme for 
this purpose.

\section{\bf Details of the Calculations}

The RHB calculations have been performed for nuclei in the neighbourhood
of the waiting-point Zn nuclei and for nuclei from r-process path to
the neutron drip line. Nuclei are treated within a spherically symmetric
configuration. The method of expansion of the wavefunctions into the
harmonic oscillator basis is employed \cite{Gambhir.90}. We have taken 20 
shells both for the fermionic as well as bosonic wavefunctions for the 
expansion.

A comparative study is made between the Lagrangian model with the
non-linear scalar self-coupling of the $\sigma$ meson and the Lagrangian 
with both the non-linear scalar self-coupling of $\sigma$ and the vector 
self-coupling of $\omega$ meson. For the former, we have used the force 
NL-SH which is a typical representative for this Lagrangian model. 
In a large number of studies it has been shown that the force NL-SH 
\cite{SNR.93} is able to provide a very good description of the 
ground-state properties of nuclei all over the periodic table. It has 
been found to be especially useful for exotic nuclei near the drip lines. 
Recently, the force NL3 \cite{Lala.97} has been brought out with a view
to improve the description of the giant monopole resonance. However,
as shown in ref. \cite{SFM.00} ground-state properties of nuclei 
predicted by NL3 are very similar to those of NL-SH. 

For the Lagrangian with the  non-linear vector self-coupling of $\omega$ 
meson which is in addition to the usual non-linear scalar self-coupling 
of $\sigma$ meson, we have used the forces NL-SV1 and NL-SV2. These
forces have been developed recently \cite{SFM.00,Sha.01} with a view to bring
about an improvement in the ground state properties of nuclei vis-a-vis
the scalar self-coupling. Consequently, it has been shown by 
Sharma et al. \cite{SFM.00} that the introduction of the vector self-coupling 
was a remedial measure in respect of the shell effects in nuclei 
about the stability line. In addition, the high-density equation of state
(EOS) of the nuclear matter becomes softer as compared to that
with the scalar self-coupling only. This would make the Lagrangian model
with the vector self-coupling compatible with the neutron star masses.
The detailed properties of these forces will be provided elsewhere 
\cite{Sha.01}.

\section{Results and Discussion}

In our attempt to scrutinize the shell effects, we started our investigation 
\cite{SFM.00} of the shell effects in nuclei at the stability line, 
where we examined the role of $\sigma$- and $\omega$-meson couplings on 
the shell effects in Ni and Sn isotopes. It was observed that the existing 
nuclear forces based upon the nonlinear scalar self-coupling of 
$\sigma$-meson exhibit shell effects which were stronger than 
suggested by the experimental data. In order to remedy this problem, 
the nonlinear vector self-coupling of $\omega$-meson in the RHB theory 
was introduced. As a result of this, the experimental data on the
shell effects in Ni and Sn nuclei on the stability line were reproduced
well \cite{SFM.00}.

\subsection{The Zn nuclei near N=50}

Having established the functional basis, we consider nuclei around $^{80}$Zn 
($N=50$) in the present work. The isotope $^{80}$Zn is assumed to be 
a waiting-point nucleus in the chain of r-process nucleosynthesis and
comes closest to the $\beta$-stability line for nuclei with $N=50$.
The experimental $S_{2n}$ value
for this nucleus is about 11 MeV. The nuclei $^{80}$Zn and $^{82}$Zn
being very rich in neutrons would serve to test the theoretical models
far away from the stability line. Here, we present the results on the shell
effects in the waiting-point Zn nuclei about $N=50$ using different
model Lagrangians. The  $S_{2n}$ values for the Zn isotopes obtained
from the RHB calculations are shown in Fig.~1. The experimental data 
is taken from the 1995 mass tables \cite{Audi.95}. Because the mass of the 
nucleus $^{82}$Zn is based upon the systematics, there may be a slight 
uncertainty in the $S_{2n}$ value of the nucleus $^{82}$Zn. Thus, the 
data point for $^{82}$Zn is shown in Fig.~1 enclosed within 
a diamond.

The RHB calculations [Fig.~1(a)] with NL-SH (scalar self-coupling only) 
show a very large kink at $N=50$ implying that the shell effects 
with NL-SH are much stronger than exhibited by the available data. 
In comparison, the shell gap with NL-SV2 (both the scalar and vector 
self-couplings) is reduced as compared to NL-SH. 
It is, however, still larger than the data shown. 
The reduction in the shell gap with the vector self-coupling is
consistent with that observed in Ni nuclei at the stability line
vis-a-vis the scalar self-coupling only.

We show in Fig.~1(b) the $S_{2n}$ values obtained with the other 
vector self-coupling force NL-SV1. The slope of the kink shows that the 
shell gap with NL-SV1 agrees well with the available data. 
Thus, RHB with the vector self-coupling Lagrangian NL-SV1 is able to 
reproduce the available data on the shell effects in the waiting-point 
region satisfactorily. We also show in Fig.~1(b) the non-relativistic HFB 
calculations with the Skyrme force SkP \cite{DFT.84}. The SkP results 
produce nearly a straight line passing through the $N=50$ point.
A comparison of these results with the data shows that 
the shell effects with SkP are quenched strongly. This is reminiscent of 
the behaviour of SkP as observed in ref. \cite{SFM.00} at the stability 
line, where the shell effects were found to be quenched strongly in Ni 
isotopes at $N=28$ in contrast to the experimental data. This feature of 
SkP is evidently due to its high effective mass $m^* \sim  1$ which 
increases the level density near the Fermi surface and thus suppresses 
the shell gaps significantly \cite{DFT.84}.

\subsection {Nuclei about N=82 near the r-process path}

Nuclei in the vicinity of $Z=40$ with the magic neutron number $N=82$ are
expected to lie on or close to the r-process path. We have chosen the
isotopic chains $Z=36-46$ with neutron numbers running across the magic 
number $N=82$. The r-process path is expected to pass through these nuclei 
especially those with the higher $Z$ values. Some nuclei in these
chains are close to the neutron drip-line as the neutron separation 
energy approaches a vanishing value. 

How the shell effects behave in nuclei near the r-process path is an open 
question ? This is due to the fact that the $N=82$ nuclei near the r-process 
path possess an extremely large neutron-to-proton ratio. Although a 
tremendous progress has been made in the last decade in synthesizing
very neutron-rich nuclei in the laboratory, such nuclei are still very 
far from being realized. The last such nucleus attained in this region 
in the laboratory is $^{130}$Cd ($Z=48$ and $N=82$). The mass of the 
nucleus $^{132}$Cd with $N=84$ is still not known. The knowledge of the 
latter would allow to discern the shell effects in regions far away from 
the stability line. For $N=82$ nuclei with $Z < 48$, i.e., in the 
region of Zr ($Z=40-44$), the magnitude of the nuclear isospin 
becomes increasingly large and hence the difficulty in producing these 
nuclei. In the present paper, we examine the shell effects in these 
nuclei, therefore, in the light of the data available in the other 
regions. We also study how the shell effects evolve as one approaches 
the continuum in moving from the r-process path towards the neutron 
drip-line.

Having described the available data on the waiting-point Zn nuclei
by the forces with the vector self-coupling as shown in Fig.~1, we extend
our formalism to explore the inaccessible region of the r-process path and
the neutron drip-line about $N=82$. With a view to visualize the evolution
of the shell effects with the isospin, we show our RHB results on the
$S_{2n}$ values for the isotopic chains from the higher $Z$ to 
the lower ones using the two Lagrangian models. Fig.~2 shows the 
corresponding values for the Pd ($Z=46$) and Ru ($Z=44$) nuclei across the 
magic number $N=82$. 

The results [Fig.~2] with the non-linear scalar self-coupling (NL-SH)
for the Pd and Ru isotopic chains show a shell gap at $N=82$, which is 
largest amongst all the forces shown here. With the force NL-SV2 with the 
vector self-coupling of the $\omega$-meson, the shell effects are milder 
as compared to NL-SH. This feature is similar to that observed at the 
stability line \cite{SFM.00} and also in Fig.~1. Furthermore, the results 
with the force NL-SV1 show that the shell gap at $N=82$ is reduced as 
compared to NL-SV2. This is again similar to that seen for the Zn 
isotopes. For a comparison, we also show the HFB results obtained 
with the interaction SkP. The shell gap with SkP is much reduced as 
compared to the forces with the vector self-coupling. Thus, the shell 
effects with NL-SV1 are stronger as compared to the HFB+SkP results 
shown in the figure. It is interesting to note that there is a remarkable
agreement in the results below and up to the magic number for all 
the forces except for NL-SH. The neutron separation energy from
these results for $N=82$ Pd and Ru nuclei is about 3-4 MeV. 
Accordingly, these nuclei are expected to lie well on the 
r-process path as per the conventional wisdom. 

We show the RHB results for the isotopic chains of Mo ($Z=42$) and 
Zr ($Z=40$) in Fig.~3. From the argument of the neutron separation
energy, the Mo and Zr nuclei about the magic number are also
expected to lie along the r-process path. Thus, both the Figs. 2 and 3
portray how the shell effects evolve in moving to nuclei with larger 
neutron excess. The main feature that the shell effects with NL-SH are 
strong is also displayed by Fig.~3. It is worth reminding that the force 
NL-SH was also used in an earlier work \cite{SLHR.94} to calculate 
the shell effects at $N=82$ in Zr nuclei about the drip line.
The shell effects with NL-SH were noted to be similarly strong in the 
previous results. However, the previous results with NL-SH were based 
upon the BCS pairing and there was no mechanism in place to include the 
effects of the coupling to the continuum. The qualitative behaviour 
that the shell effects with NL-SH are strong is reproduced here also 
in the present RHB calculations. The reason for this overestimation of 
the shell effects as also noted in ref. \cite{SFM.00} is that the 
properties which were considered mainly in constructing these forces such 
as NL-SH and others were the binding energies and charge radii of nuclei.
The shell effects as such were not included in such considerations 
heretofore. As shown in ref. \cite{SFM.00}, the shell effects emerge 
as an observable which should be employed to constrain the nuclear 
interaction. 

With the force NL-SV2 with the vector self-coupling of
the $\omega$-meson, the shell effects [Figs.~2-3] are milder
as compared to NL-SH for all the chains. This is similar
to that observed at the stability line \cite{SFM.00} and also in Fig.~1(a). 
The results with our benchmark force NL-SV1 show that the shell
gap at $N=82$ is reduced as compared to NL-SV2. This is again similar to 
that observed for the Zn isotopes. The $S_{2n}$ values obtained with
the HFB calculations using the Skyrme interaction SkP are also shown
for comparison. The magnitude of the kink at N=82 shows that the shell
effects with HFB+SkP are much weaker as compared to NL-SV1. The quenching 
predicted by SkP in this region is as expected. As noted earlier,
HFB+SkP is known to exhibit a strong quenching at the stability line 
\cite{SFM.00} and the same has also been observed about the waiting-point 
nucleus $^{80}$Zn as shown in Fig.~1(b) above. In comparison, the RHB 
approach with NL-SV1 reproduces the shell effects about the stability line 
\cite{SFM.00} as well as in the waiting-point region at $N=50$. 

For any given force, the shell gap at $N=82$ shows a steady decrease 
in moving from Pd ($Z=46$) to Zr ($Z=40$). As nuclei become increasingly
neutron rich, the gradual decrease in the $S_{2n}$ values and also in the 
corresponding shell gap shows the evolution of the shell effects in going
away from the r-process path towards the drip line. This reduction in the
shell gap is largest for SkP which shows intrinsically weaker shell effects. 
As seen from Fig.~3(b) for the Zr isotopes, the curve for $S_{2n}$ has 
become nearly a straight line, implying a disappearance of magicity with SkP.
This picture is at variance with that from NL-SV1. The persistence of the 
kink with NL-SV1 in Figs.~2 and 3 shows that with NL-SV1 the shell effects 
for the r-process  nuclei at $N=82$ are much stronger than with SkP. 
Thus, notwithstanding the fact that NL-SV1 is commensurate with the 
available data along the stability line and in the waiting point 
region, it can be stated that the shell effects in $N=82$ nuclei on 
the r-process path still remain stronger.

\subsection{Nuclei near the neutron drip-line}

Nuclei become increasingly unbound and a coupling to the continuum arises 
in going to extremely large neutron to proton ratios. This situation is 
expected to arise naturally in going to $N=82$ nuclei below $Z=40$. For this
situation, we show the results for Sr ($Z=38$) and Kr ($Z=36$) isotopes in 
Fig.~4. The results show that the $S_{2n}$ value for  the $N=82$ nuclei 
decreases from $\sim$ 2.5 MeV for Sr isotopes to $\sim$  1.5 MeV. 
The corresponding $S_n$ value would amount to a merely 1 MeV or less. 
On the basis of the $S_n$, the $N=82$ Kr nucleus represents the arrival 
of the drip line.

The shell effects with NL-SV1 are seen to become successively weaker 
(which is true for all the forces) as one moves to nuclei with higher 
isospin such as $^{120}$Sr [Fig.~4(a)] and $^{118}$Kr [Fig.~4(b)]. 
Here, the case of the Kr isotopes [Fig.~4(b)] deserves a mention. 
Except with NL-SH, all the other forces show that the shell effects
are washed out significantly. This stems from the fact that for $^{118}$Kr
the Fermi energy is very close to the continuum and the nucleus is 
pushed to the very limit of binding. The binding energy of an additional 
neutron is then close to zero and the shell gap ceases to exist. 
Thus, any semblance of the shell effects for $^{118}$Kr ($N=82$) is 
completely lost. This situation is destined to befall on a nucleus or
nuclei for every magic number, leading to a termination of the 
corresponding shell gap. All the forces NL-SV2, NL-SV1 and SkP 
predict such a behaviour except NL-SH. This situation at the drip 
line at the magic number can be termed as {\it terminus stratum}.

The nuclei close to the drip line show no binding to an extra neutron 
to be able to form a higher mass isotope. Such nuclei at the 
drip line ($\lambda_n \sim 0$) would not contribute to the r-process
nucleosynthesis. In view of this, it is therefore important to understand 
the nature of the shell effects for nuclei along the r-process path, 
which are assumed to play an important role in the nucleosynthesis,
rather than those at the drip line.

\subsection{The single-particle levels}

We show in Fig.~5 the neutron single-particle levels obtained with
NL-SV1 for $N=80$ nuclei from a few chains. Our focus is the evolution 
of the shell gap at $N=82$ as one approaches the drip line 
($\lambda_n \sim 0$) from the r-process path. As discussed in the 
subsection above, the Mo and Zr nuclei are expected to lie on the
r-process path, whereas the Sr and Kr nuclei should be very close to the 
drip line. The single-particle levels shown in this figure portray 
essentially the results of Figs.~3-4. The Fermi energy (shown by the dashed 
lines) approaches the continuum as one moves towards a larger neutron to 
proton ratio. The shell gap ($N=82$) shows a constant decrease as the last 
neutrons become more and more unbound in going from Mo to Kr. 
For $^{118}$Kr ($N=82$) the Fermi energy is close to zero and the 
shell gap merges into the continuum. However, for nuclei on the 
r-process path such as Mo and Zr shell gaps do remain large.
For the corresponding nuclei of Pd and Ru, the shell gaps are
(not shown here) even larger than that for Mo. 

The difference in the response of the various forces near the drip line is 
illustrated in the single-particle levels shown for $^{120}$Zr in Fig.~6. 
Since this nucleus has 2 neutron less than the magic numbers, all
the occupied levels are bound. Moreover, we do not see any indication
of a partial occupancy of any level above the Fermi energy, which might
arise out of the Bogoliubov pairing. The figure shows that the $N=82$ 
shell gap is largest with NL-SH and it shows a reduction in going to NL-SV1.  
This is consistent with the general behaviour of the shell effects from 
NL-SH to NL-SV1 as seen in Figs.~2-4. On the other hand, the shell gap 
with SkP is reduced significantly as compared to NL-SV1. The reduced shell 
strength with SkP seems to be generic as shown above. 

In Fig.~7 we have shown the neutron single-particle levels for the 
nucleus $^{124}$Zr with 2 neutrons more than the magic number $N=82$. 
Here we have made a comparison of the levels for NL-SH, NL-SV1 and SkP.
The shell gap at $N=82$ shows a decreasing trend from NL-SH to SkP as
seen also in Fig.~6. However, it can be noticed that the extra 
2 neutrons above the magic number go to the levels in the continuum and
therefore with the force NL-SH and NL-SV1 this nucleus is unbound.
This should imply that $^{124}$Zr is already a drip line nucleus, whereas 
the same is not the case for $^{120}$Zr (see Fig.~6). The  single-particle 
levels with SkP show a remarkable difference from those of the RMF forces. 
With SkP the $2f_{7/2}$ level is below the continuum and the extra 2 
neutrons occupy this level predominantly. Thus, the nucleus $^{124}$Zr 
is bound within the HFB+SkP calculations in contrast to the RMF forces. 

\subsection{The spin-orbit potential}

The spin-orbit potential is the key ingredient of the nuclear interaction
responsible for creation of the magic numbers and the shell gaps. As
mentioned earlier, the spin-orbit interaction arises naturally in the
RMF theory as a result of the Dirac-Lorentz structure of nucleons.
The $\sigma$ and $\omega$ potentials contribute constructively to 
generate the required spin-orbit strength \cite{Serot.86}. This
is able to reproduce the experimental data on spin-orbit splittings 
in nuclei such as $^{16}$O and $^{40}$Ca. It has been shown by several 
authors \cite{Rein.89,SNR.94} that the effective mass in the 
RMF theory can be constrained by the spin-orbit splittings in light 
nuclei to a value close to 0.60. However, the shell gaps and thus the 
shell effects in heavier nuclei can not be addressed adequately by 
fixing the spin-orbit splittings in $^{16}$O only. This becomes clear
by examining the successful RMF forces  such as NL-SH and NL3 which
show stronger shell effects as compared to the experimental data 
in heavy nuclei \cite{SFM.00}. This implies that the shell effects 
in heavier nuclei need to be taken into account in order to make
a successful nuclear interaction. In this respect, the experimental
data on $S_{2n}$ values around the shell closure provide an additional
useful ground-state observable.

The advantage which the RMF theory derives from its intrinsic spin-orbit
interaction is not enjoyed by the non-relativistic approaches such
as the Hartree-Fock Skyrme theories. The spin-orbit potential is
usually added on an ad-hoc basis and the strength of the spin-orbit
potential is fitted in order to reproduce the spin-orbit splittings 
in light nuclei in these approaches. This may, however, not suffice 
for the shell effects in heavy nuclei as illustrated for the case of 
the RMF theory. 

The spin-orbit (s.o.) potential plays a central role in determining the 
magnitude of the spin-orbit splittings in nuclei. We show in Fig.~8
the spin-orbit potential obtained in the RHB theory for the Lagrangian
parameters NL-SH, NL-SV2 and NL-SV1. The contribution of the $\sigma$ and
$\omega$ terms add together to give rise to the spin-orbit potential
which is predominant in the surface region of the nucleus \cite{Serot.86}. 
The s.o. potential is largest for the
non-linear scalar force NL-SH. This is consistent with the strong
shell effects observed with NL-SH at the shell closures in nuclei.
The depth of the s.o. potential is reduced by about 10 \% for the force
NL-SV2. It is reduced further with NL-SV1. This reduction from NL-SH to
NL-SV2 and to NL-SV1 is consistent with all the results we have
discussed on the shell effects with these forces in this work. This
underlines the point that the spin-orbit interaction plays a key role
in determining the strength of the shell effects. This would require a
an adjustment of both the $\sigma$ and the $\omega$ fields in order to 
produce a suitable spin-orbit field.

\section{Conclusion}

We have investigated the evolution of the shell effects in nuclei 
from the r-process path to the neutron drip line. The theoretical
approach that is employed in the present work is that of the
Relativistic Hartree-Bogoliubov theory based upon the quasi-particle
scheme with the self-consistent pairing. The RHB approach with the
inclusion of the vector self-coupling of $\omega$ meson was developed
with a view to remedy the problem of the strong shell effects
with the Lagrangian with the scalar self-coupling only. Having reproduced
the shell effects with the vector self-coupling at the stability line
in our earlier work \cite{SFM.00}, we have attempted to reproduce the 
available data on the shell effects in the waiting-point region of $^{80}$Zn.
The available data on Zn isotopes are reproduced well within the RHB 
approach with the vector self-coupling. This establishes further the 
reliability of the working basis we employ.

With the above approach we have carried out the RHB calculations for 
nuclei along the r-process path and near the neutron drip-line using 
the two different Lagrangian models. It is shown that the shell 
effects with the vector self-coupling are in general softer than those 
with the scalar self-coupling. This is found to be the case over a 
broad range of nuclei considered in our work. This implies that the 
shell structure near the Fermi surface with the vector self-coupling 
forces is denser than with the Lagrangian with the scalar self-coupling. 
This is consistent with what was found to be the case also for nuclei 
about the stability line.

The shell gap at $N=82$ shows a gradual decrease in going from the r-process 
path towards the neutron drip line for all the forces considered. This is
the natural behaviour for the shell gaps to evolve with the
isospin. However, we show that for nuclei on the r-process path, 
the shell effects at $N=82$ still remain strong vis-a-vis a quenching 
exhibited by HFB+SkP. This quenching with SkP has been discussed
extensively in the literature \cite{Kratz.93,DHNA.94,Chen.95}. 
The origin of the shell quenching stems primarily from the very large 
effective mass of SkP which leads to a significant compression of the 
shell structure near the Fermi surface. A quenching has been requested 
for an improved fit to the global r-process abundances 
\cite{Kratz.93,Chen.95}. However, a quenching does not seem not to be
compatible with the data at the stability line and in the waiting-point 
region.

In moving away from the r-process path where nuclei are expected
to have $S_n \sim 2-4$ MeV to nuclei where $S_n$ is vanishingly low,
the shell gap at N=82 merges into the continuum. Consequently, 
the ensuing shell structure above N=82 is washed out significantly. 
Since a nucleus can not take on any more neutrons, nuclei in the vicinity 
of the drip line are, therefore, not expected to contribute to the 
r-process nucleosynthesis.

This work is supported by the Research Administration Project No. SP056 
of the Kuwait University. We thank Karlheinz Langanke for fruitful 
discussions.




\addvspace{16mm}

\newpage
\begin{itemize}

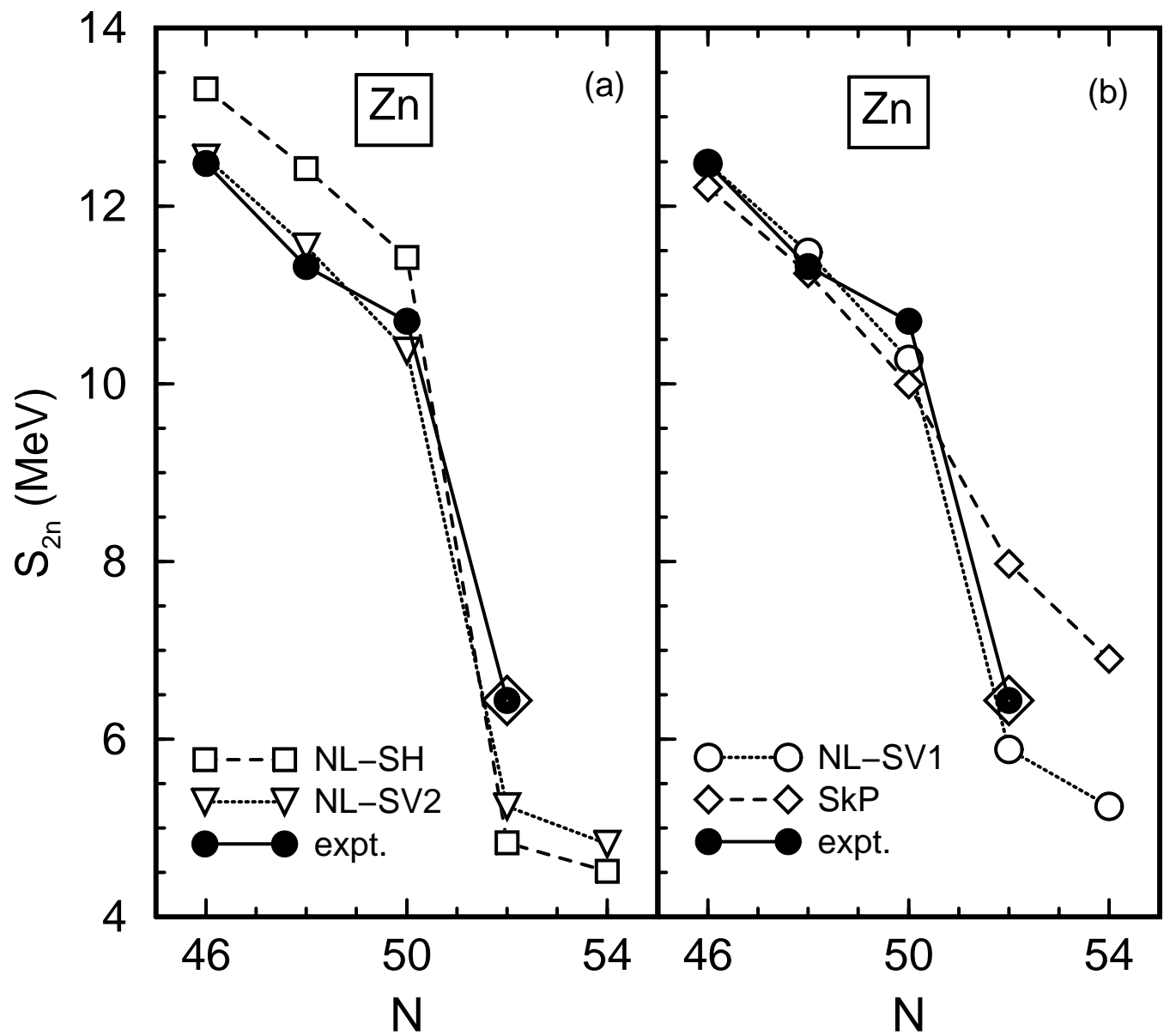
\item[Fig. 1] The $S_{2n}$ values for Zn isotopes in the waiting-point 
region obtained with RHB using (a) the scalar self-coupling force 
NL-SH and the vector self-coupling force NL-SV2 and (b) with the vector 
self-coupling force NL-SV1 along with the nonrelativistic HFB+SkP results. 
The experimental data is shown for comparison. The datum on $^{82}$Zn
is adopted from the mass systematics \cite{Audi.95} and is shown enclosed 
within a diamond. 

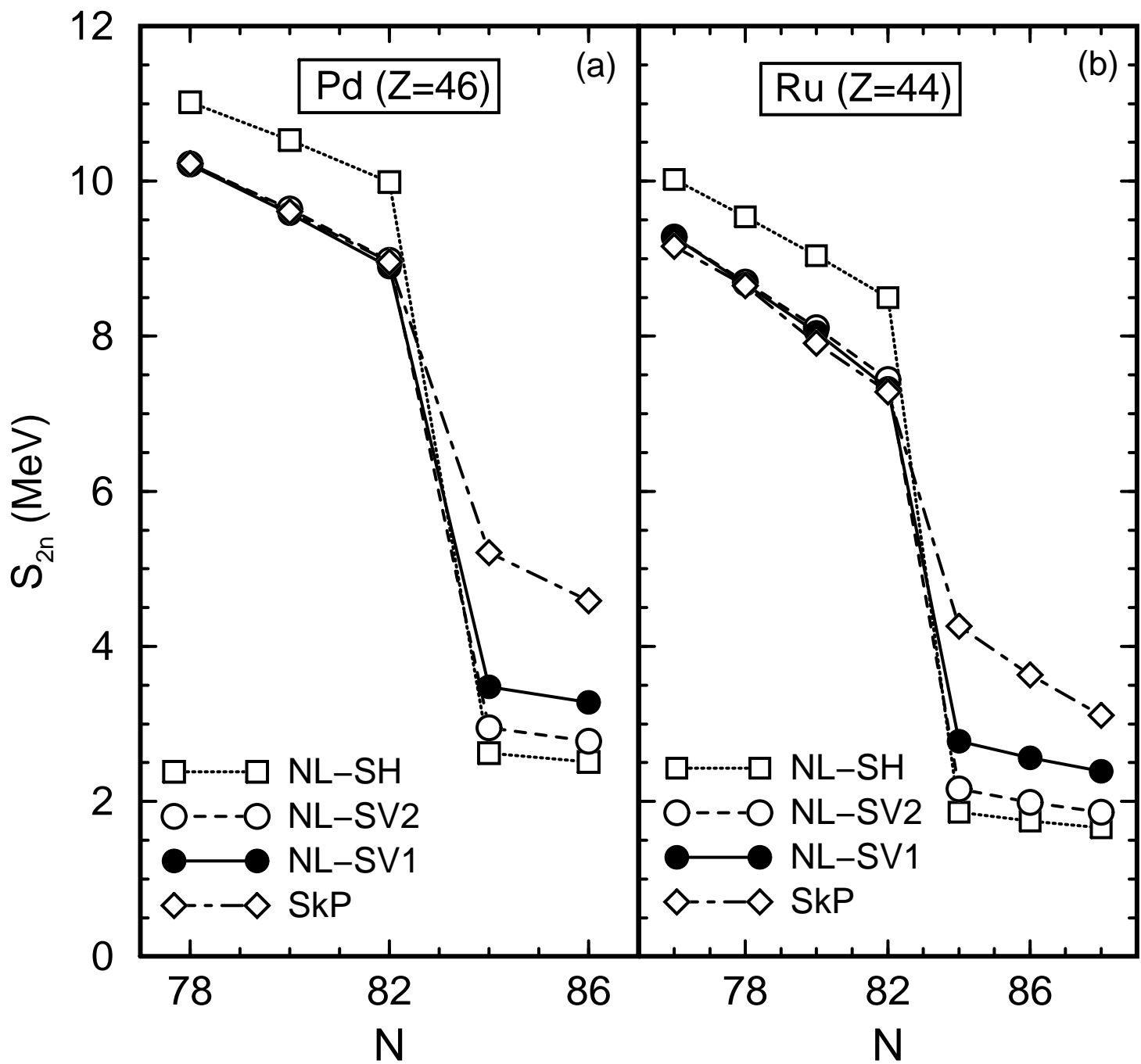
\item[Fig. 2] The RHB results on the shell effects at $N=82$ in the r-process
nuclei of (a) Pd and (b) Ru. The HFB+SkP results are also shown for a
comparison.

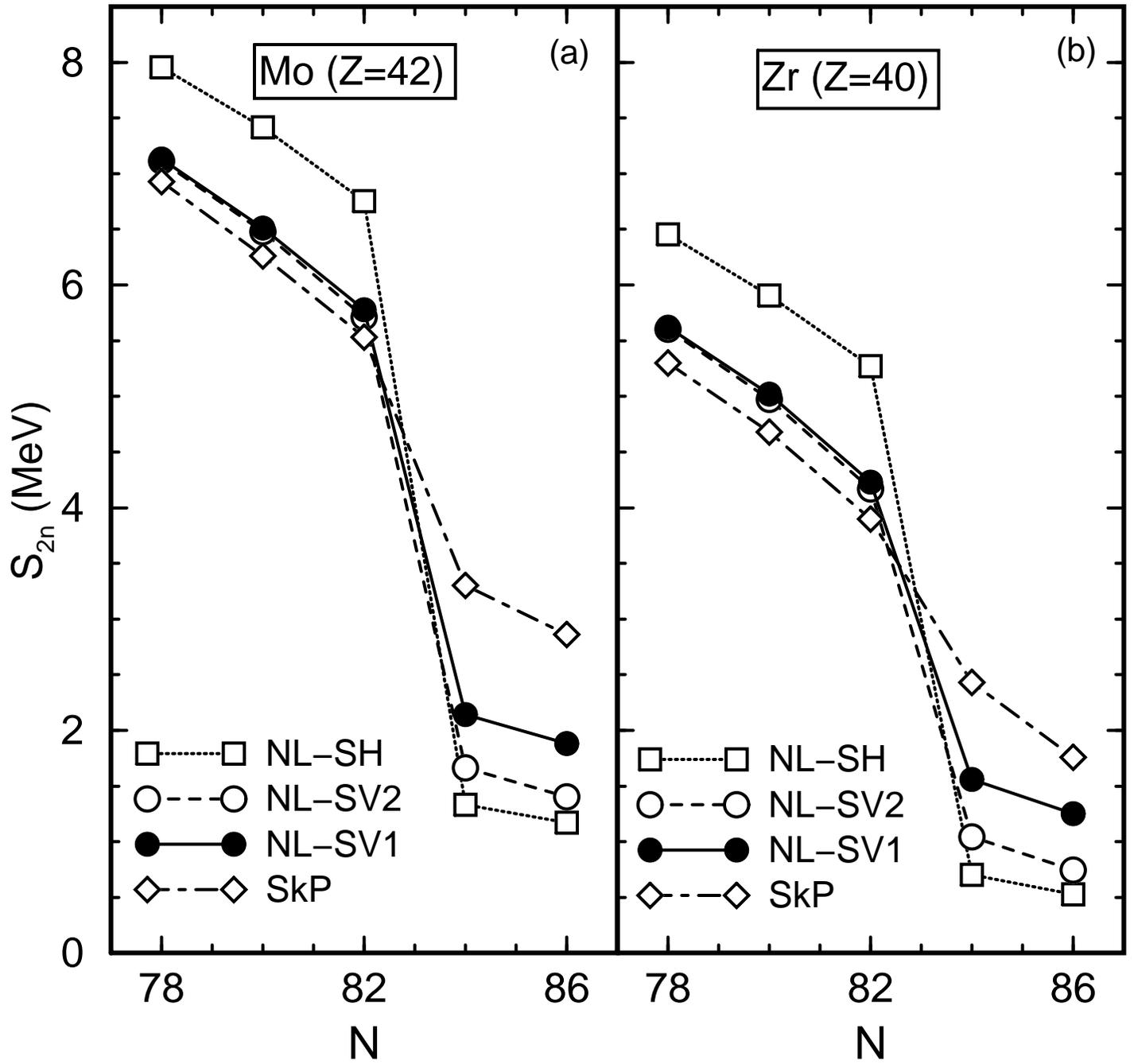
\item[Fig. 3] The same as Fig.~2 for the r-process nuclei (a) Mo and
(b) Zr. The $S_{2n}$ values and shell gaps show a gradual decline in moving
from Pd to Zr.

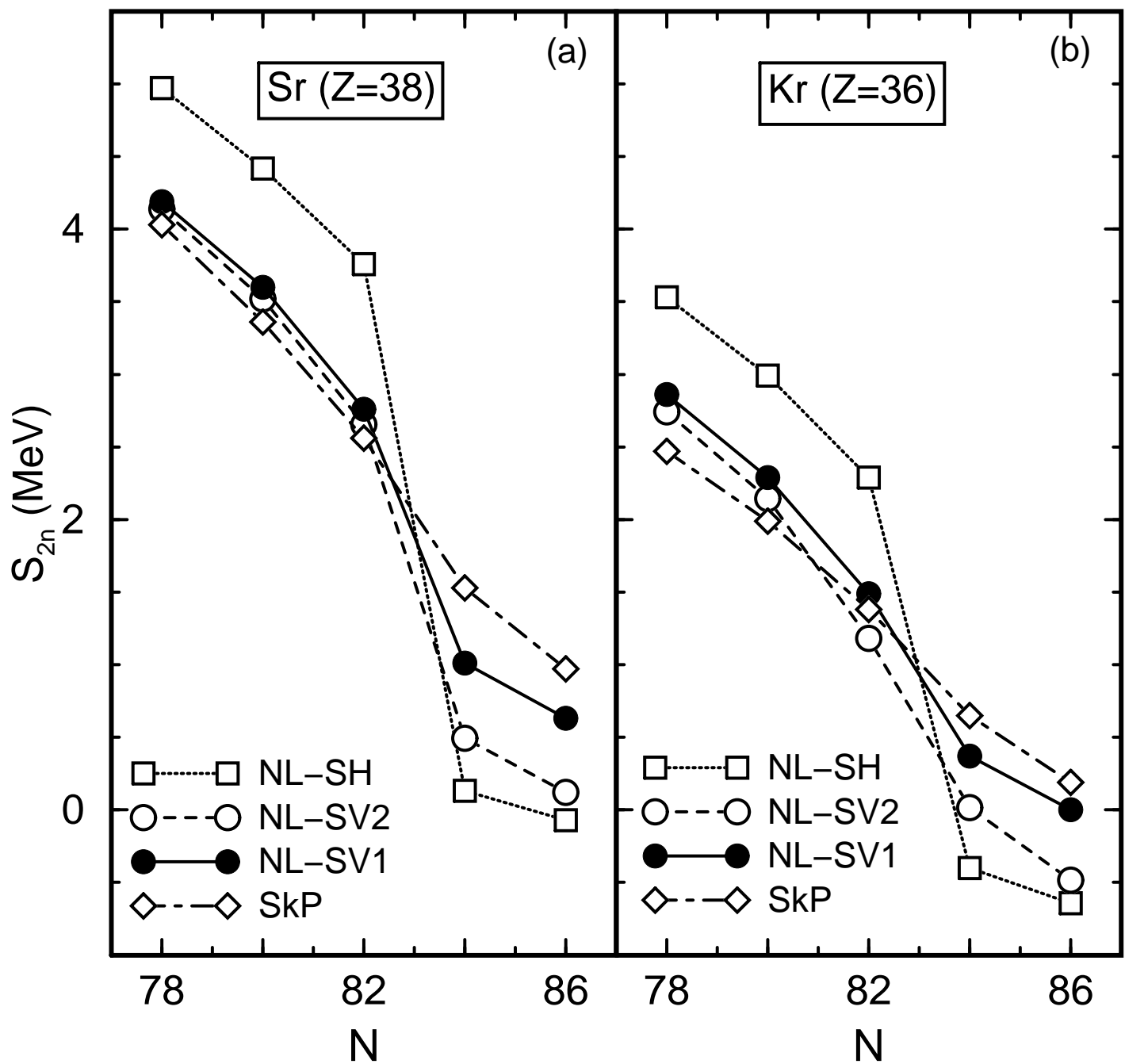
\item[Fig. 4] The same as Fig.~2 for the drip line nuclei near $N=82$
for the Sr and Kr isotopic chains. The shell gap and the $S_{2n}$ values 
show a vanishing trend for the Kr isotopes about $N=82$.

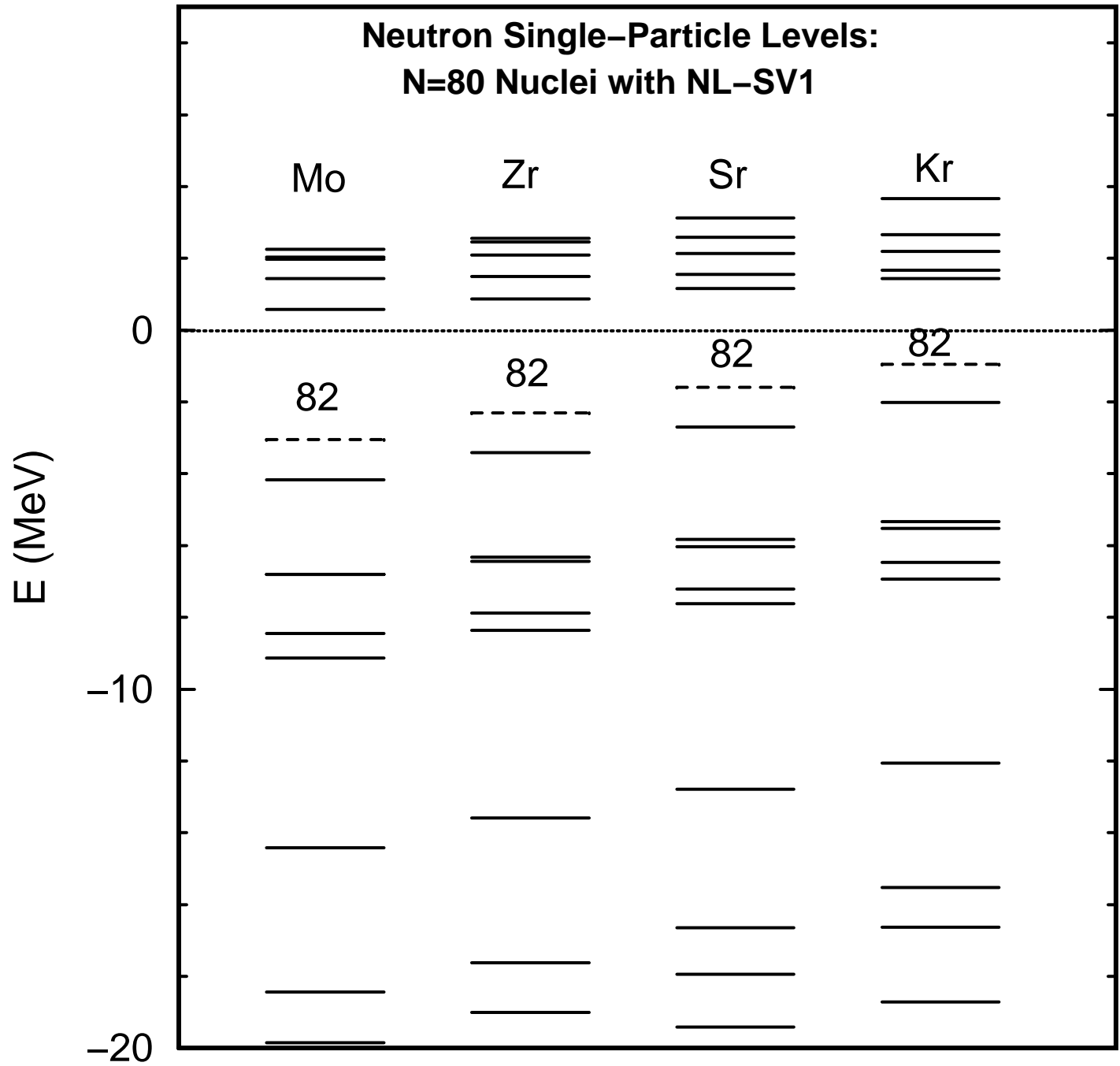
\item[Fig. 5] The neutron single-particle levels for the Mo, Zr, Sr and 
Kr nuclei ($N=80$) showing an evolution of the shell gap in moving
from the r-process path towards the drip line, with the force NL-SV1. 
The Fermi energy is shown by the dashed lines. The Fermi energy for the 
Kr nucleus near the drip line is seen to merge with the continuum.

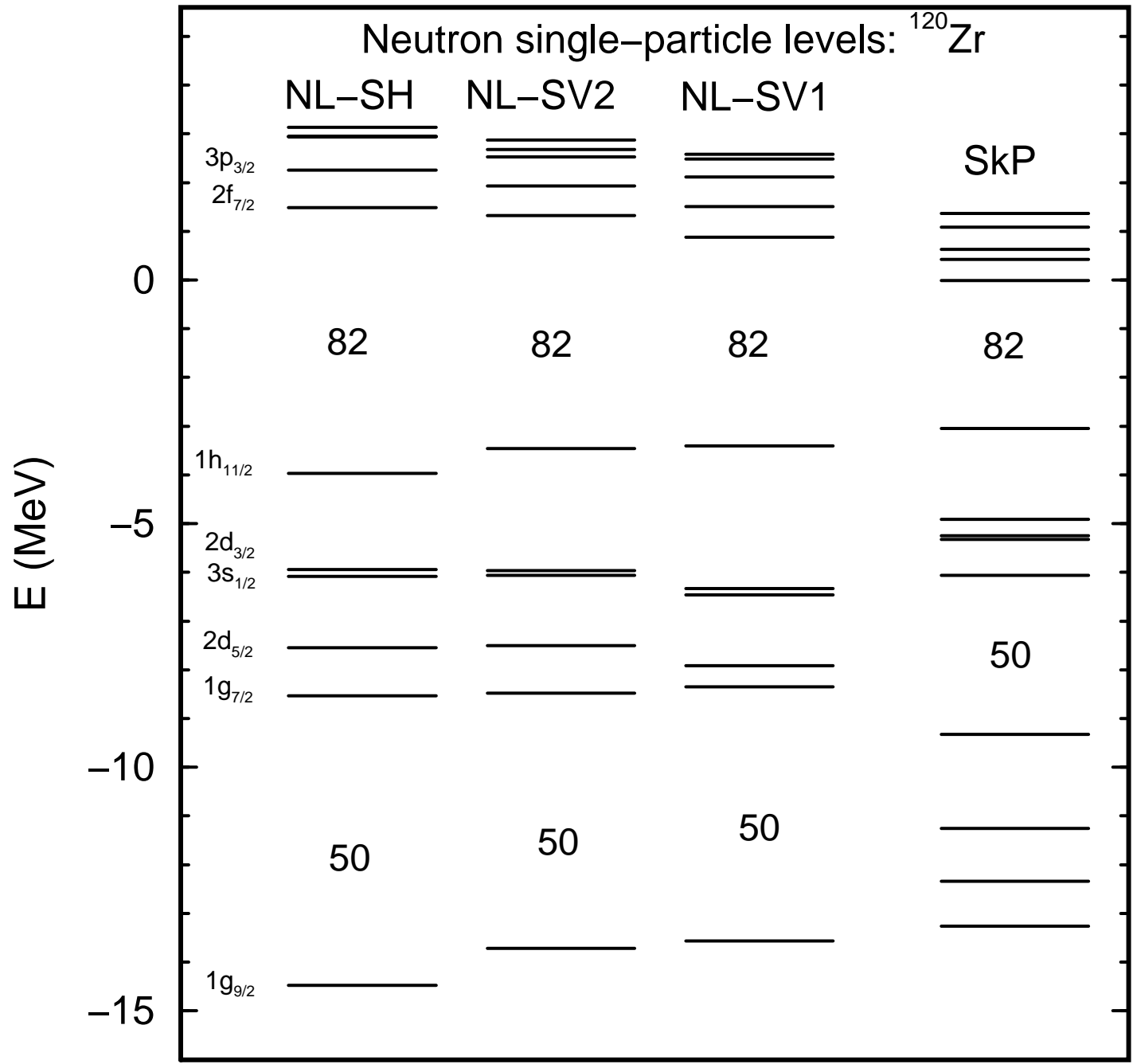
\item[Fig. 6] The shell gap $N=82$ for the nucleus  $^{120}$Zr using
the various interactions.

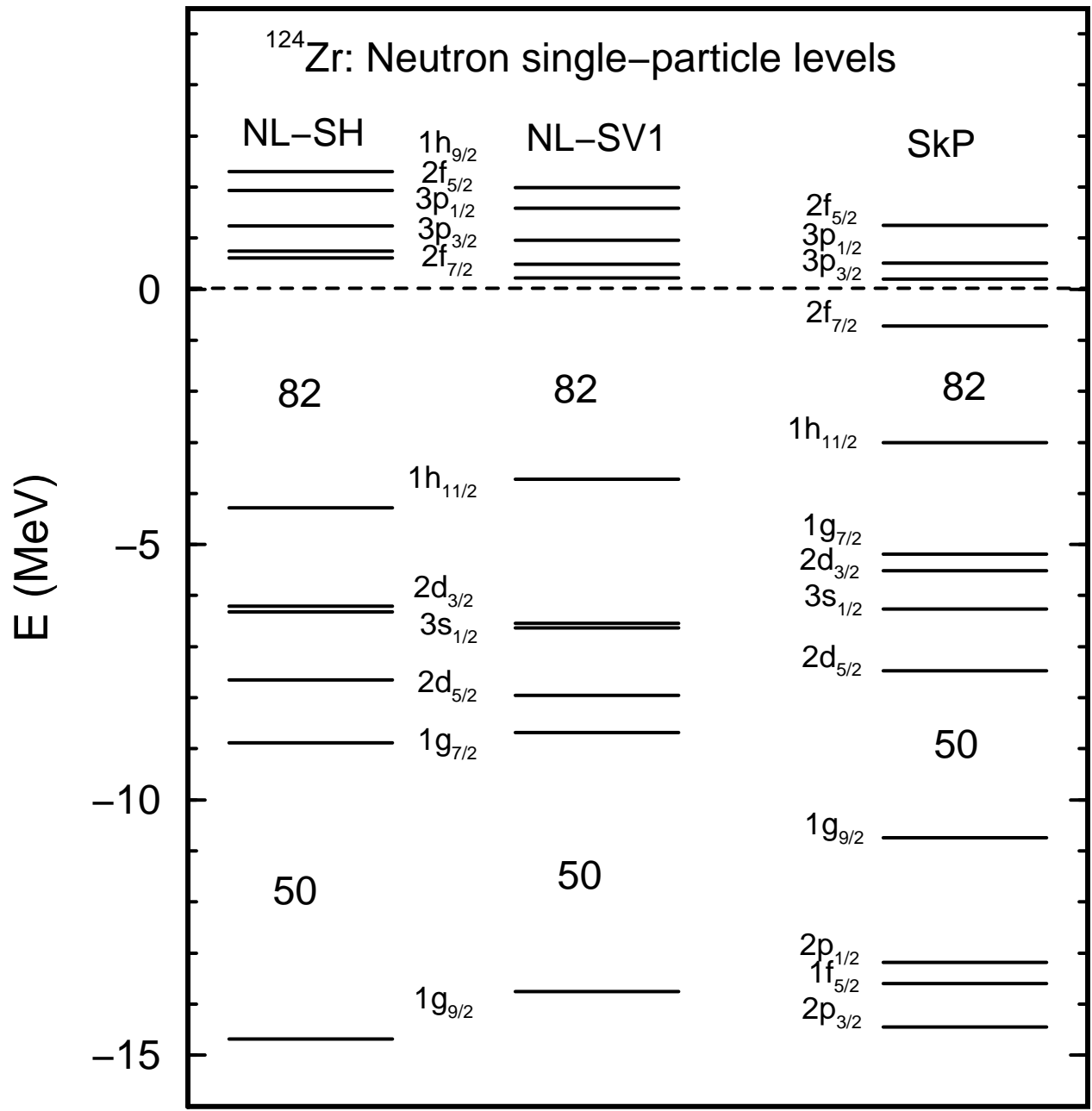
\item[Fig. 7] The neutron single-particle levels for the nucleus
$^{124}$Zr with 2 neutrons above the closed shell. The reduction
of the $N=82$ shell gap from NL-SH to NL-SV1 to SkP is clearly evident.
In contrast to the RMF forces, the $2f_{7/2}$ level is bound with
SkP. 

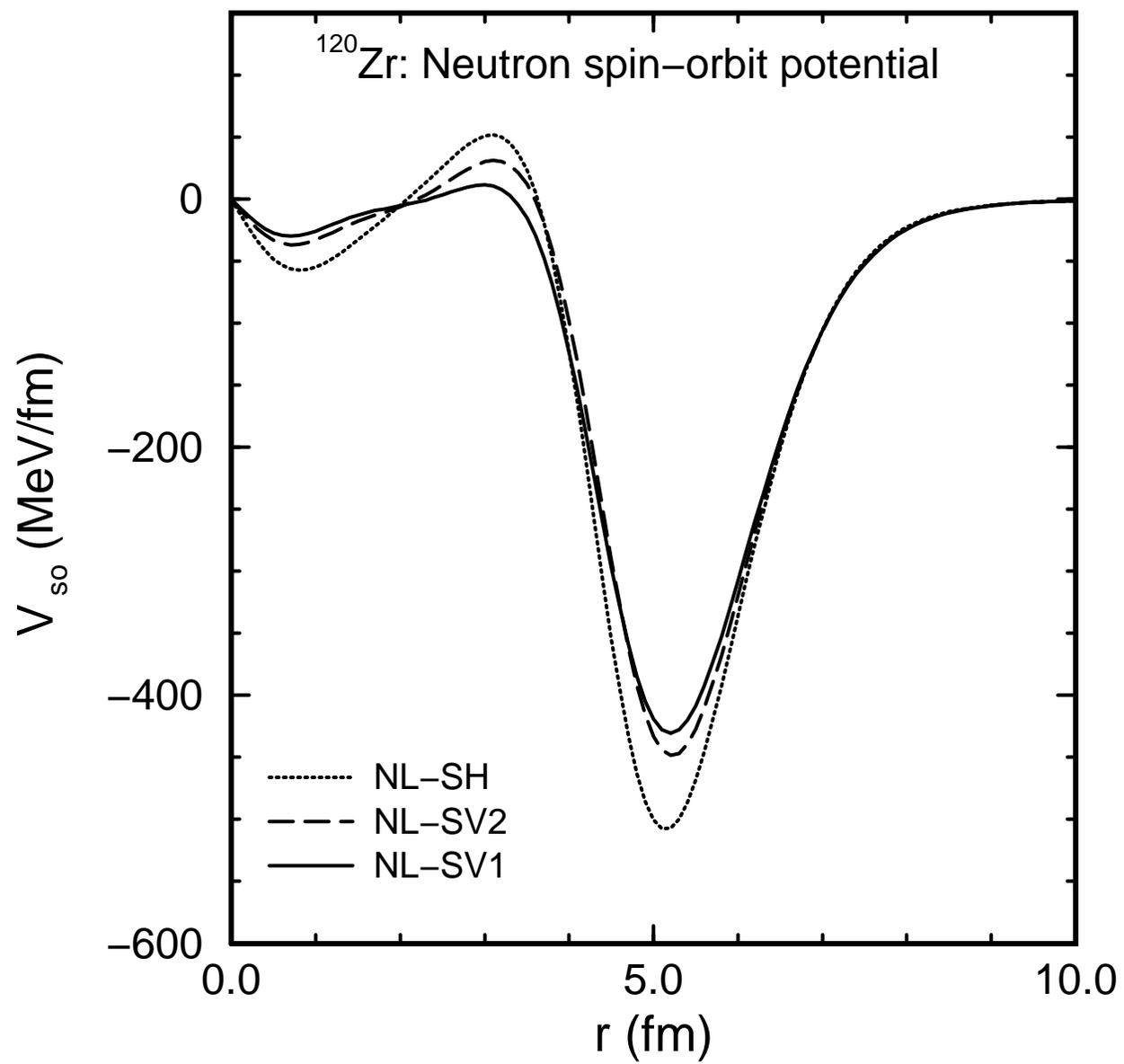
\item[Fig. 8] The neutron spin-orbit potential for $^{120}$Zn obtained
in the RHB calculations with the various RMF forces. The predominance of
the spin-orbit potential in the surface region is clearly evident. 

\end{itemize}


\end{document}